\documentclass[]{aa}
\usepackage{graphicx}
\usepackage[varg]{txfonts}
\usepackage{natbib}
\usepackage{balance}
\bibpunct{(}{)}{;}{a}{}{,} % to follow the A&A style
\usepackage[figuresright]{rotating}
\usepackage[a4paper,breaklinks]{hyperref}

\def\kms{$\mathrm{km\, s^{-1}}$}

\newcommand{\logg}{\ensuremath{\log g}}

\newcommand{\mlp}{\ensuremath{\alpha_{\mathrm{MLT}}}}

\newcommand{\draftflag}{false}

\newcommand{\beq}{\begin{equation}}
\newcommand{\eeq}{\end{equation}}

\newcommand{\COBOLD}{{\sf CO$^5$BOLD}}

\newcommand{\cobold}{\COBOLD}

\newcommand{\xx}{\ensuremath{\mathrm{1D}_{\mathrm{LHD}}}}

\idline{10}{10}

\begin{document}

\title{X-Shooter GTO: Chemical analysis of a sample of EMP candidates\thanks{Based
on observations obtained at ESO Paranal Observatory,
GTO programme 086.D-0094}
}

\author{
E.~Caffau\thanks{Gliese Fellow}\inst{1,2} \and
P. Bonifacio,  \inst{2} \and
P. Fran\c cois,\inst{3,2} \and
M. Spite,      \inst{2} \and
F. Spite,      \inst{2} \and
S. Zaggia,     \inst{4} \and
H.-G. Ludwig,  \inst{1,2} \and
L. Monaco,     \inst{5} \and
L.~Sbordone,   \inst{1,2,6} \and
R. Cayrel,     \inst{2} \and
F. Hammer,     \inst{2} \and
S. Randich,    \inst{7} \and
V. Hill,       \inst{8} \and
P. Molaro      \inst{9}
}

\institute{ 
Zentrum f\"ur Astronomie der Universit\"at Heidelberg, Landessternwarte, 
K\"onigstuhl 12, 69117 Heidelberg, Germany
\and
GEPI, Observatoire de Paris, CNRS, Univ. Paris Diderot, Place
Jules Janssen, 92190
Meudon, France
\and
UPJV, Universit\'e de Picardie Jules Verne, 33 Rue St Leu, F-80080 Amiens
\and
Istituto Nazionale di Astrofisica,
Osservatorio Astronomico di Padova Vicolo dell'Osservatorio 5, 35122 Padova, Italy
\and
European Southern Observatory, Casilla 19001, Santiago, Chile
\and
Max-Planck Institut f\"ur Astrophysik, Karl-Schwarzschild-Str. 1, 
85741 Garching, Germany
\and
Istituto Nazionale di Astrofisica,
Osservatorio Astrofisico di Arcetri, Largo E. Fermi 5, 50125 Firenze, Italy
\and
Universit\'e de Nice Sophia Antipolis, CNRS,
Observatoire de la C\^ote d'Azur, Laboratoire Cassiop\'e e, B.P. 4229, 
06304 Nice Cedex 4, France
\and
Istituto Nazionale di Astrofisica,
Osservatorio Astronomico di Trieste,  Via Tiepolo 11,
I-34143 Trieste, Italy
}
\authorrunning{Caffau et al.}
\titlerunning{EMP stars}
\offprints{E.~Caffau}
\date{Received ...; Accepted ...}

\abstract%
%\context
{
Extremely metal-poor stars (EMP) are very rare
objects that hold in their atmospheres the fossil
record of the chemical composition of the early
phases of Galactic evolution.
Finding these objects and determining their
chemical composition provides important constraints on these
early phases.
}
%\aims
{Using a carefully designed selection method,
we chose a sample of candidate EMP stars from
the low resolution spectra of the Sloan Digital Sky Survey
and observed them with X-Shooter at the VLT to
confirm their metallicities and determine abundances for as many
elements as possible.}
%\method
{The X-Shooter spectra are analysed by means of one-dimensional,
plane-parallel, hydrostatic model atmospheres. Corrections
for the granulation effects are computed using CO5BOLD
hydrodynamical simulations.}
%\results
{All the candidates are confirmed to be EMP stars, proving the
efficiency of our selection method within about 0.5\,dex.
The chemical
composition of this sample is  
compatible with those of brighter samples, suggesting that the
stars in the Galactic halo are well mixed.
}
%\conclusions 
{These observations show that it is feasible
to observe, in a limited amount of time, a large sample of
about one hundred stars among EMP candidates selected from the SDSS.
Such a size of sample will allow us, in particular, to confirm
or refute the existence of a vertical
drop in the Galactic Halo metallicity distribution
function around [Fe/H]$\sim -3.5$.
}

\maketitle

%%%%%%%%%%%%INTRODUCTION%%%%%%%%%%%%%%%%%%%%%%%%%%%%%%%%%%%

\section{Introduction}

Stellar data show that the Galaxy, after the Big Bang, was very
rapidly enriched in elements heavier than Li by the
ejecta of the first stars. The stars observable nowadays are more or
less metal-rich because they formed from material more or less enriched in
heavy elements.  
Low mass stars (${\rm M}<0.8{\rm M}_\odot$), have mean-sequence lifetimes 
older than the age of the Universe and, if they formed at the beginning
of the Universe, 
before significant metal enrichment took place,
they would still shine today and could be observable. 
It is nevertheless
difficult to find these rare extremely metal-poor stars, which are the
relics of the earliest phases of the Galactic evolution. Dedicated
surveys have been designed (the most relevant being
\citealt{beers85,beers92,beers99} and \citealt{christlieb08} but the
earlier surveys of \citealt{Bond70,SB71,BM73}, are also relevant), with
some striking \citep{BN84,molaro90,MB90,christlieb,frebel,norris}, 
but limited successes, owing to the rarity of these stars.  
The number of extremely metal-poor stars is small, and their study has
raised some questions, because in many of them the carbon abundance exceeds
by more than two orders of magnitude the value expected if they had
a solar-scaled composition \citep{christlieb,frebel,norris}: as
a consequence, the global metallicity of these objects is comparable
to that of a typical metal-poor star of about [Fe/H]=-2.5.

One way to extend the sample is to use surveys of greater sky
coverage, at the cost of handling large quantities of data.  Another
solution is to extend the surveys to fainter magnitudes, which means the
exploration of a larger volume, finding more remote faint targets.
With a strategy of combining adapted sophisticated criteria for target
selection and high-efficiency spectrographs for the high-resolution
follow-up (X-Shooter), a few more extremely metal-poor stars should be
found.  This is what we tried with the SDSS data for the target
selection, and we present here the results.  We took the opportunity
of the French-Italian X-Shooter guaranteed time observations (GTOs) to
observe a sample of good candidates of turn-off, extremely metal-poor
(EMP) stars. The high efficiency of X-Shooter permits us to
observe in about one hour a 16-17\,mag star, during one night between
four and eight stars can be observed.

%%%%%%%%%%%%%%%%%%%%%%%%%%%%%%%%%%%%%%%%%%%%%%%%%%%%%%%%%%%%%%%%%%

\section{Target selection}

The Sloan Digital Sky Survey \citep[SDSS,][]{sdss} archive is a real
treasure chest when searching for rare objects, which we have exploited
in the past five years in our quest to identify extremely metal-poor 
candidates. The survey is
optimised to observe galaxies, but many stars are in the sample of
observed objects. SDSS provides photometry in five broad bands
specifically tied to extragalactic surveys: $u$ centred at 355.1\,nm,
$g$ at 468.6\,nm, $r$ at 615.5\,nm, $i$ at 748.1\,nm, and $z$ at
893.1\,nm.  We used the $(g-z)_0$ colour to derive the effective
temperature of the stars \citep{ludwig08}. For a subsample of objects,
a wide range spectrum (380-920\,nm) is provided, at low resolution
(R=1800-2000), and low signal-to-noise ratio (hereafter S/N) of 
about 20 at 400\,nm for a
$g=17$ object.  We selected objects classified as stars in the SDSS
archive, with both photometric and spectroscopic observations
available and turn-off star colours. In this way, we were able to fix
the gravity at 4.0 on a logarithmic scale (c.g.s. units).  The sample
of about 125\,000 spectra obtained from this selection was analysed
with an automatic procedure, to derive the metallicity of the
stars. We inspected by eye all the spectra that were found to be more
metal-poor than $\left[{\rm Fe/H}\right]=-2.5$, in order to select the
most promising candidates.  Eight of these candidates were
observed during the French-Italian guaranteed time observation (GTO)
of the X-Shooter spectrograph at VLT.  Two of them were observed
in April and May 2010 \citep{bonifacio11}, and six in February 2011.
One of the candidates observed in the last run appears particularly
interesting, showing no iron line in the X-Shooter spectrum.
A dedicated paper on this star is in preparation.
For the selection of the candidates presented in this work, taking
into account the target coordinates, only about 40\,000 objects, out
of the sample of 125\,000, were observable, and 754 of them have a
metallicity that is lower than $\left[{\rm Fe/H}\right]\le -3.5$.
These numbers tell us that extremely-metal-poor stars might not be so
rare as generally assumed, and that a systematic programme to observe
them could significantly extend the number of these primitive stars
detected.

%%%%%%%%%%%%%%%%%%%%%%%%%%%%%%%%%%%%%%%%%%%%%%%%%%%%%%%%%%%%%%%%%%
\section{Observations and data reduction}

The observations were performed on the night of 10 February 2011
with Kueyen (VLT UT2) and the high efficiency spectrograph X-Shooter
\citep{dodorico}. In Table \ref{allstar}, we present the coordinates and some
basic photometric informations about the programme stars. 
The log of the observation is presented in Table
\ref{logbook}.  The X-Shooter spectra range from 300\,nm to 2400\,nm
and gathered by three detectors.  The observations have been performed in
staring mode with $1\times2$ on-chip binning along the spectral
direction and with the integral field unit (IFU), which re-images an
input field of 4"x 1.8" into a pseudo slit of 12"x 0\farcs{6}
\citep{IFU}.  As no spatial information was reachable for our targets,
the aim of using the IFU was to use it as a slicer with three
0\farcs{6} slices.  This corresponds to a resolution of R=7900 in the
UVB arm and R=12600 in the VIS arm.  The spectra were reduced using
the X-Shooter pipeline \citep{goldoni}, which performs the bias and
background subtraction, cosmic ray hit removal \citep{vandokkum01},
sky subtraction \citep{kelson03}, flat-fielding, order extraction, and
merging. However, the spectra were not reduced using the IFU pipeline
recipes. Each of the three slices of the spectra were instead reduced
separately in slit mode with a manual localisation of the source and
the sky. This method allowed us to perform an optimal extraction of the
spectra leading to an efficient cleaning of the remaining cosmic ray
hits but also to a noticeable improvement in the S/N.

{\small}
\begin{table*}
\caption{\label{allstar}
Coordinates and photometric data.}
\tabskip=0pt
\begin{center}
\begin{tabular}{lccrrrrrrrrr}
\hline\noalign{\smallskip}
\multicolumn{1}{l}{SDSS ID}& 
\multicolumn{1}{c}{RA}&
\multicolumn{1}{c}{Dec}& 
\multicolumn{1}{c}{$u$}& 
\multicolumn{1}{c}{$g$}& 
\multicolumn{1}{c}{$r$}&
\multicolumn{1}{c}{$i$}&
\multicolumn{1}{c}{$z$}&
\multicolumn{1}{c}{J} &
\multicolumn{1}{c}{H} &
\multicolumn{1}{c}{K} &
\multicolumn{1}{c}{A$_{V}$}\\
 &  J2000.0 & J2000.0 & [mag] & [mag] & [mag] & [mag] & [mag] & [mag] & [mag] & [mag]  \\
\noalign{\smallskip}\hline\noalign{\smallskip}
J044638-065529 & 04\, 46\, 38.234&$-$06\, 55\, 29.07 & $19.454$ & $18.592$ & $18.224$ & $18.102$ & $18.001$ & $-$ & $-$ & $-$ & $0.354$ \\
J082511+163459 & 08\, 25\, 11.458 & +16\, 34\, 59.97 & $18.622$ & $17.563$ & $17.089$ & $16.875$ & $16.774$ & $16.048$ & $15.349$ & $15.359$ & $0.142$ \\
J085211+033945 & 08\, 52\, 11.519 & +03\, 39\, 45.15 & $17.834$ & $16.906$ & $16.646$ & $16.534$ & $16.492$ & $15.765$ & $15.301$ & $15.355$ & $0.184$ \\
J090733+024608 & 09\, 07\, 33.285 & +02\, 46\, 08.17 & $17.277$ & $16.329$ & $16.010$ & $15.869$ & $15.826$ & $15.005$ & $14.642$ & $14.784$ & $0.095$ \\
J133718+074536 & 13\, 37\, 18.760 & +07\, 45\, 36.31 & $19.181$ & $18.277$ & $18.041$ & $17.957$ & $17.923$ & $-$ & $-$ & $-$ & $0.118$ \\
\noalign{\smallskip}\hline\noalign{\smallskip}
\end{tabular}
\end{center}
Optical magnitudes are from SDSS, infrared magnitudes from 2MASS.
\end{table*}

\begin{table*}
\caption{\label{logbook}
Log of the observations. 
}
\begin{tabular}{lrrrrl}
\hline\noalign{\smallskip}
\multicolumn{1}{c}{Star}& 
\multicolumn{1}{c}{date}&
\multicolumn{1}{c}{Exp Time (sec)}& 
\multicolumn{1}{c}{mode}&
\multicolumn{1}{c}{S/N @ 650 nm }&
\multicolumn{1}{l}{SDSS Obj. Type$^a$}
\\ 
\noalign{\smallskip}\hline\noalign{\smallskip}
SDSS\,J044638-065528  &  2011-02-12 &  5400   &    IFU  &  110 & Ser BLUE \\
SDSS\,J082511+163459  &  2011-02-12 &  5400   &    IFU  &  130 & RED STD \\
SDSS\,J085211+033945  &  2011-02-12 &  3600   &    IFU  &  150 & SP STD \\
SDSS\,J090733+024608  &  2011-02-12 &  1800   &    IFU  &  120 & SP STD \\ 
SDSS\,J133718+074536  &  2011-02-12 &  3450   &    IFU  &   80 & RED STD  \\
\noalign{\smallskip}\hline\noalign{\smallskip}
\multicolumn{6}{l}{
$^a$ Object type, based on colours that motivated the selection of this
object for SDSS spectroscopy:}\\
\multicolumn{6}{l}{
SP STD = spectrophotometric standard;
Ser BLUE = serendipity blue object; RED STD = reddening standard.}
\end{tabular}
\end{table*}

The use of the IFU can cause some problems with the sky subtraction because
there is only $\pm 1$\,arcsec on both sides of the object.
In the case of a large gradient in the spectral flux (caused by
emission lines), the modelling of the sky background signal
can be of poor quality owing to the small
number of points used in the modelling.

%%%%%%%%%%%%%%%%%%%%%%%%%%%%%%%%%%%%%%%%%%%%%%%%%%%%%%%%%%%%%%%%%%
\section{Model atmospheres}

The analysis was done with 1D, plane-parallel, hydrostatic model
atmospheres, computed in local thermodynamical equilibrium (hereafter LTE) 
with ATLAS\,9 \citep{kurucz93,kurucz05} in its
Linux version \citep{sbordone04,sbordone05}, line
opacity being treated using opacity distribution functions (ODFs). We used the
ODFs computed by \citet{castelli03}, with a microturbulent velocity of
1\kms. Convection was treated in the mixing length approximation
with \mlp=1.25 and the overshooting option in ATLAS was switched off.
For each star, we computed an ATLAS\,9 model, with the parameters
derived from the photometry. For the abundance corrections related to
thermal inhomogeneities and differences in stratification
(3D corrections hereafter), we used the
3D-\cobold\ model \citep{freytag02,freytag10} with the closest
parameters in the CIFIST grid \citep[see][]{ludwig09} and the
associated \xx\ model \citep{cl07} as reference.

%%%%%%%%%%%%%%%%%%%%%%%%%%%%%%%%%%%%%%%%%%%%%%%%%%%%%%%%%%%%%%%%%%
\section{Analysis}

The effective temperature was derived from the photometry ($g-z$
colour), by using the calibration described in \citet{ludwig08}.  We
also fitted the wings of the H$\alpha$ line with a grid of synthetic
profiles computed using ATLAS models and a modified version of the
{\tt BALMER} code\footnote{The original version is available on-line
  at \url{http://kurucz.harvard.edu/}}, which uses the theory of
\citet{barklem00,barklem00b} for the self-broadening and the profiles
of \citet{stehle99} for Stark broadening.  For three stars of our
sample (SDSS\,J044638-065528, SDSS\,J085211+033945, and
SDSS\,J133718+074536), we find a good consistency of ${\rm T_{\rm eff}}$ 
from the H$\alpha$ and photometry, to within 100\,K.
SDSS\,J090733+024608 displays an H$\alpha$ fitting temperature cooler than the
photometric one by about 300K, while SDSS\,J082511+163459 is 500\,K
cooler. In this latter star, the red wing of H$\alpha$ in the observed
spectrum is affected by an absorption that could explain, at least in
part, the disagreement between the effective temperature derived from the
H$\alpha$ fitting profile and from photometry.  We decided to keep a
coherent effective temperature and our choice was the photometric one
because we consider this to be more robust.  X-Shooter is an echelle
spectrograph, and the continuum placement in the H$\alpha$ region,
which greatly affects the temperature determination, cannot be easily
performed in an objective way. For the gravity, we assumed for all the
stars that $\logg = 4.0$, because the stars are selected from the colours
to be at the turn-off. X-Shooter spectra do not provide us with many
possibilities to derive the gravity. For three stars, some \ion{Fe}{ii}
lines are detectable, and the comparison of the iron abundance from
the two ionization states can be used as a check of the gravity. For
one of them (SDSS\,J082511+163459) the agreement is good, considering
to the quality of the data (0.11\,dex). For SDSS\,J085211+033945 and
SDSS\,J090733+024608, the difference is non-negligible, 0.33 and
0.27\,dex, respectively, but the two \ion{Fe}{ii} lines visible in the
X-Shooter spectra are very weak and give a difference in abundance of
0.15 and 0.14\,dex, respectively.

\begin{figure}
\begin{center}
\resizebox{\hsize}{!}{\includegraphics[draft = \draftflag,clip=true]
{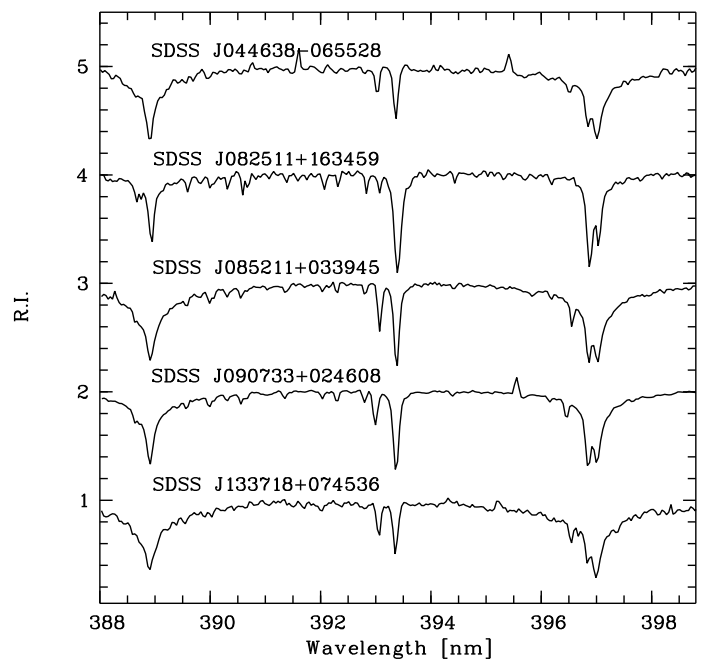}}
\end{center}
\caption[]{The X-Shooter observed spectra, in the range of the 
K and H-\ion{Ca}{ii} lines for the five stars of the sample.
}
\label{obscaii}
\end{figure}

Owing to the ``low'' resolution of the X-Shooter spectra we were unable to
derive the abundance from weak lines, nor
derive the micro-turbulence directly from the abundance versus (vs.)
equivalent width (EW)
relation, which we derived instead using the formula in \citet{edvardsson93}.
A change in the microturbulence of $\Delta\xi =0.5$\kms\ induces a change
in the iron abundance of $\Delta\left[{\rm Fe/H}\right]\sim
0.15$\,dex. For the same reason, we were unable to
detect several elements.  The solar abundances that we used was this from
\citet{abbosun} for Fe, and those from \citet{lodders09} for the other elements. 
The atomic parameters adopted for the spectral-synthesis fitting
are from the ``First Stars'' survey
\citep{bonifacio09,cayrel04,francois}.
In Fig.\,\ref{obscaii}, the observed spectra of the five stars of the
sample are shown in the range of the K and H-\ion{Ca}{ii} lines.

%%% FIGURE %%%%%%%%%%%%%%%%
%%% FIGURE %%%%%%%%%%%%%%%%

Owing to the low resolution, several lines are blended
and some lines cover only a few pixels. We therefore
derived the abundance by fitting the observed lines against a grid
of synthetic profiles, rather than measuring the equivalent widths. We used
a modified version of the code developed in \citet{bonifacio03}. We
used a grid of ATLAS+SYNTHE synthetic spectra in which only the
abundance of the element measured was varied and minimised the $\chi ^2$ of
the difference between  observed and synthetic spectrum, where the abundance of the
element, continuum placement, and line shift were free parameters.  We
first derived the iron abundance, then considered this as a fixed value in the
computation of the synthetic grid for the other elements.  

A summary of the stellar parameters and the main results of the
sample of stars is available in Table\,\ref{analysis}.
Tables\,\ref{ab044}-\ref{ab133718} 
contain the abundances of the elements that can be studied
in the X-Shooter spectra of the five stars of our sample. 
The listed uncertainties are the line-to-line scatter,
and are consistent with the S/N of the observed spectra. For only
one star in the sample were we able to derive the Li abundance from the Li
doublet at 670.7\,nm. The upper limits and the A(Li) determination can
be found in Table\,\ref{analysis}.

\begin{table*}
\caption{\label{analysis}
The stellar parameters and main results.
}
\begin{center}
\begin{tabular}{lrrrrrrrrr}
\hline\noalign{\smallskip}
Star                 & ${\rm V_{\rm rad}}$ & ${\rm T_{\rm eff}}$ & \logg & $\xi$ &S/N & [Fe/H]$_{\rm SDSS}$ & [Fe/H] & [$\alpha$/H] & A(Li) \\
                    & \kms & \kms & K                 & \kms  &  @ 400\,nm &      &    \\
\hline\noalign{\smallskip}
SDSS\,J044638-065528 & 242 & 6194 & 4.0 & 2.0 &  45 & $-3.38$ & $-3.71\pm 0.27$ & $-3.12\pm 0.11$ & $<2.4$\\
SDSS\,J082511+163459 &  24 & 5463 & 4.0 & 1.5 &  90 & $-3.58$ & $-3.22\pm 0.24$ & $-3.18\pm 0.16$ & $<1.4$\\
SDSS\,J085211+033945 & 228 & 6343 & 4.0 & 2.0 &  90 & $-3.15$ & $-3.24\pm 0.24$ & $-2.86\pm 0.22$ & $<2.5$\\
SDSS\,J090733+024608 & 304 & 5934 & 4.0 & 1.8 & 100 & $-3.37$ & $-3.52\pm 0.14$ & $-3.11\pm 0.13$ & $2.2$\\
SDSS\,J133718+074536 & 212 & 6377 & 4.0 & 2.0 &  40 & $-4.40$ & $-3.49\pm 0.32$ & $-3.27\pm 0.14$ & $<2.8$\\
\noalign{\smallskip}\hline\noalign{\smallskip}
\end{tabular}
\end{center}
\end{table*}

%%%%%%%%%%%%%%%%%%%%%%%%%%%%%%%%%%%%%%%%%%%%%%%%%%%%%%%%%%%%%%%%%%
\subsection{SDSS\,J044638-065528}

Only the Fe, Mg, and Ca abundances can be derived from the X-Shooter
spectrum.  Five lines of \ion{Mg}{i} are detectable in the
spectrum. Two features (382.9, 383.2\,nm) lie on the wings of H lines,
but with the line profile fitting technique we can take this into
account.  The other three lines (the Mg-b triplet) are in a noisy
region of the spectrum, and the bluest line, at 516.7\,nm, is blended
with an iron line.  For this reason, only the line at 518.3\,nm of the
triplet is kept for the abundance determination. From these selected
three Mg lines, magnesium is enhanced with respect to iron by about
+0.6\,dex.  Four Ca lines are present in the spectrum.  The only
\ion{Ca}{i} line at 422.6\,nm is noise dominated, hence we
cannot rely on it. The \ion{Ca}{ii}-K line gives $\left[{\rm
Ca/H}\right]=-3.44$.  Fitting simultaneously the two reddest lines
of the IR \ion{Ca}{ii} triplet, we obtain $\left[{\rm
    Ca/H}\right]=-2.60$, while two separate fits provide $-2.72$ and
$-2.30$, respectively.  The sky subtraction was problematic 
which might explain the disagreement on the Ca abundances 
derived from the different Ca lines.
Deviations from LTE could be another
reason.  The abundances derived for all elements are in
Table\,\ref{ab044}.  No feature is clearly detectable in the
wavelength domain where the Li doublet at 670.7\,nm would be observed,
and the upper limit is given in Table\,\ref{analysis}.

\begin{table}
\caption{\label{ab044}
SDSS\,J044638-065528
abundances.
}
\begin{center}
\begin{tabular}{llr}
\hline\noalign{\smallskip}
Element & [X/H]$_{\rm fit}$ &  N\\
       & x=2.0\kms         &   \\
\hline\noalign{\smallskip}
\ion{Fe}{i}  & $-3.71\pm 0.27$ & 20\\
\ion{Mg}{i}  & $-3.11\pm 0.11$ & 3\\
\ion{Ca}{i}  & $-3.95 $        & 1\\
\ion{Ca}{ii} & $-2.51\pm 0.30$ & 2\\
\noalign{\smallskip}\hline\noalign{\smallskip}
\end{tabular}
\end{center}
\end{table}

The 3D corrections play a non-negligible role in the abundance
determination in metal-poor stars. For this star, the 3D correction
\citep{cl07,abbosun} applied to iron would reduce [Fe/H] by $-0.18$\,dex 
($\left[{\rm Fe/H}\right]_{\rm 3D}=-3.89\pm 0.35$).  
For Mg, the 3D correction is smaller in absolute value and positive 
at +0.10\,dex.  For the line
of \ion{Ca}{i}, it is $-0.13$\,dex, while for it is \ion{Ca}{ii}
$-0.23$\,dex.

%%%%%%%%%%%%%%%%%%%%%%%%%%%%%%%%%%%%%%%%%%%%%%%%%%%%%%%%%%%%%%%%%%
\subsection{SDSS\,J082511+163459}

Lines of several elements are detectable in the spectrum of this star
(see Table\,\ref{ab0825}), which is one of the two most metal-rich
stars in this sample.  Two lines of \ion{Fe}{ii} are visible in the
spectrum, and provide an abundance in very close agreement with the one
derived from the lines of neutral iron. Several lines of
$\alpha$-elements can be detected but their pattern is not clear,
hence they do not all give the same $\alpha$-enhancement. From the lines
available, Mg is not enhanced with respect to Fe but we excluded from
the abundance determination the line at 516.7\,nm, which is blended with an
iron line.  Silicon is slightly enhanced, but only one line is
detectable in the observed spectrum. From four lines of \ion{Ti}{ii},
an enhancement of about 0.5\,dex is derived. Lines of neutral and singly
ionised Ca are detectable in the spectrum, but provide discrepant results: 
two \ion{Ca}{i} lines imply that there is a slight enhancement of Ca, similar to
that seen for Si, while four lines of \ion{Ca}{ii} indicate
that Ca behaves in a similar way to Ti. This disagreement might imply
that the gravity is not correct, but the two ionisation states of iron
give very similar iron abundances,
possibly because the \ion{Ca}{ii}-K line is very strong (EW
of about 157\,pm) and difficult to model precisely; and we know that
the IR-triplet is in a spectral range where sky subtraction is
problematic.  Finally, deviations from LTE could also be a source of
the discrepancy.  In any case, owing to the limited spectral resolution and S/N, the
uncertainty in the abundances is about 0.2\,dex. We averaged the
abundance derived from three lines of Mg and the Ca lines, excluding
the IR \ion{Ca}{ii} triplet, to derive the enhancement of the
$\alpha$-elements. From the \ion{Sr}{ii} line at 407.7\,nm, a low Sr
abundance is derived, but this is not exceptional. In
\citet{francois}, the Sr abundance as a function of iron abundance
has a large scatter.  
Nickel is enhanced relatively to iron by
0.25\,dex, but this is not unusual because other metal-poor stars display a similar
behaviour \citep{sdss-uves}.  The individual abundances derived for
all elements can be found in Table\,\ref{ab0825}.

\begin{table}
\caption{\label{ab0825}
SDSS\,J082511+163459
abundances.
}
\begin{center}
\begin{tabular}{llr}
\hline\noalign{\smallskip}
Element & [X/H]$_{\rm fit}$ &  N\\
        & x=1.5\kms         &   \\
\hline\noalign{\smallskip}
\ion{Fe}{i}  & $-3.22\pm 0.24$ & 36\\
\ion{Fe}{ii} & $-3.11\pm 0.01$ & 2\\
\ion{Mg}{i}  & $-3.27\pm 0.11$ & 3\\
\ion{Al}{i}  & $-4.27\pm 0.19$ & 2\\
\ion{Si}{i}  & $-3.04$         & 1\\
\ion{Ca}{i}  & $-3.08\pm 0.22$ & 2\\
\ion{Ca}{ii} & $-2.75\pm 0.26$ & 4\\
\ion{Ti}{ii} & $-2.68\pm 0.05$ & 4\\
\ion{Cr}{i}  & $-3.38$         & 1\\
\ion{Co}{i}  & $-2.28$         & 1\\
\ion{Ni}{i}  & $-2.95\pm 0.21$ & 5\\
\ion{Sr}{ii} & $-3.71$         & 1\\
\noalign{\smallskip}\hline\noalign{\smallskip}
\end{tabular}
\end{center}
\end{table}

\begin{figure}
\begin{center}
\resizebox{\hsize}{!}{\includegraphics[draft = \draftflag,clip=true]
{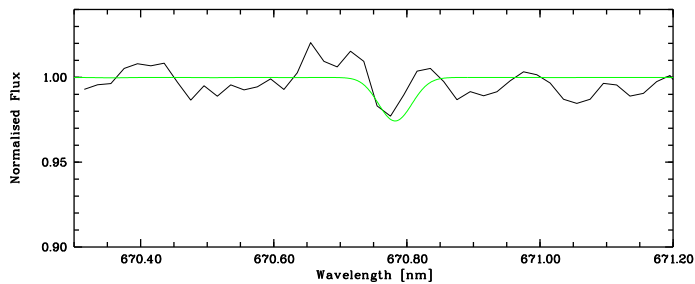}}
\end{center}
\caption[]{The range of the Li doublet for the star SDSS\,J082511+163459:
the observed spectrum (solid black) is compared to a synthetic profile
with A(Li)=1.4 (solid green/grey).
}
\label{plot_lisdssj825}
\end{figure}

The Li feature is not clearly detectable. 
A feature at the approximate wavelength 
is compatible with
a Li abundance of 1.4 (see Fig.\,\ref{plot_lisdssj825}), a value 
not anomalous for stars of this temperature.

%%% FIGURE %%%%%%%%%%%%%%%%
%%% FIGURE %%%%%%%%%%%%%%%%

%%%%%%%%%%%%%%%%%%%%%%%%%%%%%%%%%%%%%%%%%%%%%%%%%%%%%%%%%%%%%%%%%%
\subsection{SDSS\,J085211+033945}

This star, with SDSS\,J082511+163459, is the most metal-rich of the
sample, but being about 900\,K hotter than the previous one, its
spectrum contains fewer detectable lines, and the abundances of fewer
elements can be derived. At variance with SDSS\,J082511+163459,
this star has a clear enhancement in
$\alpha$-elements.

We did not use the \ion{Ca}{ii} IR triplet, which
suffers from the aforementioned sky subtraction problems, affecting
in particular the two bluest lines. 
The $\alpha$ element abundance, [$\alpha$/H]$=-2.86\pm
0.22$ is based on Mg and Ca lines.  One single line of \ion{Si}{i}
indicates that the Si abundance is in agreement with the other
$\alpha$ elements. From seven lines of \ion{Ti}{ii}, the Ti abundance
is found to be about 0.2\,dex higher than the stated
[$\alpha$/H], with a comparable $\sigma$ of 0.24\,dex.  The detection
of two \ion{Al}{i} lines permits us to derive the Al abundance. The
two lines of \ion{Sr}{ii} are also visible.  The abundances of the
elements that can be derived from the observed spectra are in Table\,\ref{ab0852}.
There is an emission feature at the place of the Li 670.8 nm doublet.

\begin{table}
\caption{\label{ab0852}
SDSS\,J085211+033945
abundances.
}
\begin{center}
\begin{tabular}{llr}
\hline\noalign{\smallskip}
Element & [X/H]$_{\rm fit}$ &  N\\
       & x=2.0\kms         &   \\
\hline\noalign{\smallskip}
\ion{Fe}{i}  & $-3.24\pm 0.24$ & 32\\
\ion{Fe}{ii} & $-2.91\pm 0.15$ & 2\\
\ion{Mg}{i}  & $-2.92\pm 0.21$ & 6\\
\ion{Al}{i}  & $-3.77\pm 0.06$ & 2\\
\ion{Si}{i}  & $-3.08$         & 1\\
\ion{Ca}{i}  & $-2.85\pm 0.20$ & 2\\
\ion{Ca}{ii} & $-2.53$         & 1\\
\ion{Ti}{ii} & $-2.63\pm 0.24$ & 7\\
\ion{Sr}{ii} & $-3.04\pm 0.01$ & 2\\
\noalign{\smallskip}\hline\noalign{\smallskip}
\end{tabular}
\end{center}
\end{table}

%%%%%%%%%%%%%%%%%%%%%%%%%%%%%%%%%%%%%%%%%%%%%%%%%%%%%%%%%%%%%%%%%%
\subsection{SDSS\,J090733+024608}

There are two \ion{Fe}{ii} lines, which provide an iron abundance higher
by 0.27\,dex than the one derived from the 37 \ion{Fe}{i} lines.

\begin{figure}
\begin{center}
\resizebox{\hsize}{!}{\includegraphics[draft = \draftflag,clip=true]
{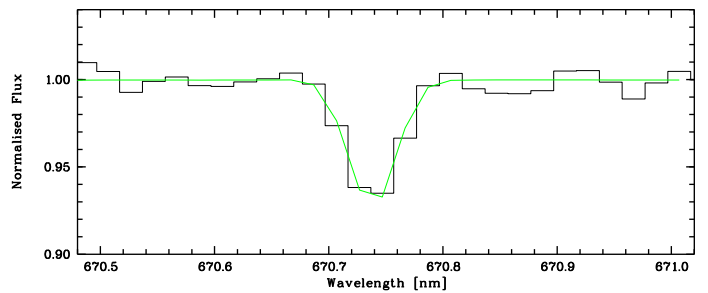}}
\end{center}
\caption[]{The range of the Li doublet for the star SDSS\,J090733+024608:
the observed spectrum (solid black) is compared to a synthetic profile
(solid green/grey). The Li abundance derived is ${\rm A(Li)_{\rm 3D-NLTE}}=2.2$.
}
\label{li0907}
\end{figure}

Among the six features of Mg detected in the observed spectrum, we retain
only four. The bluest line of the \ion{Mg}{i}-b triplet, blended with
an iron line, very often implies higher abundances than the
other two lines of the triplet. We also reject the line at 383.2\,nm,
which is located on the wing of an H line. For the calcium abundance, we rely
on the \ion{Ca}{i} lines at 422.6\,nm and on the \ion{Ca}{ii}-K line.
The IR triplet is also detectable, and for this star the sky
subtraction is not a problem, but the Ca abundance is much higher
([Ca/H]$=-2.40\pm 0.08$). In this case, we may also suspect deviations
from LTE to be the source of the discrepancy.  For the
[$\alpha$/H]$=-3.11\pm 0.13$ determination, we selected the Mg lines
and the Ca lines used for the abundance determination. The Ti
abundance, based on five \ion{Ti}{ii} lines, is about 0.3\,dex higher
than the [$\alpha$/H] value.  The Cr, Sr, and Ni abundances are each based
on one single weak line.  Both Sr and Ba
are measurable. While Sr is comparable to Fe such that [Sr/Fe]=+0.06, Ba is
enhanced by 0.3\,dex with respect to iron. The ratio $\left[{\rm
Sr/Ba}\right]=-0.36$ corresponds to a pure r-process ratio
\citep[see e.g.][]{QW}. A decrease in temperature does not change
this ratio, while a change of $\Delta\xi =\pm 0.5$\kms\ results in a
change $\Delta\left[{\rm Sr/Ba}\right] =^{-0.42}_{-0.26}$. In any
case, the Ba abundance is higher than expected. The individual
abundances are listed in Table\,\ref{ab0907}.  The Li feature at
670.7\,nm is clearly visible, and has an EW of 4\,pm meaning an
abundance compatible with the Spite plateau, ${\rm A(Li)_{\rm
3D-NLTE}}=2.2$ (see Fig.\,\ref{li0907}).

%%% FIGURE %%%%%%%%%%%%%%%%
%%% FIGURE %%%%%%%%%%%%%%%%

\begin{table}
\caption{\label{ab0907}
SDSS\,J090733+024608
abundances.
}
\begin{center}
\begin{tabular}{llr}
\hline\noalign{\smallskip}
Element & [X/H]$_{\rm fit}$ &  N\\
       & x=1.8\kms         &   \\
\hline\noalign{\smallskip}
\ion{Fe}{i}  & $-3.52\pm 0.16$ & 37\\
\ion{Fe}{ii} & $-3.25\pm 0.14$ & 2\\
\ion{Mg}{i}  & $-3.14\pm 0.14$ & 4\\
\ion{Al}{i}  & $-4.09\pm 0.06$ & 2\\
\ion{Ca}{i}  & $-3.12$         & 1\\
\ion{Ca}{ii} & $-2.98$         & 1\\
\ion{Ti}{ii} & $-2.81\pm 0.09$ & 5\\
\ion{Sc}{ii} & $-3.14$         & 1\\
\ion{Cr}{i}  & $-3.79$         & 1\\
\ion{Ni}{i}  & $-3.31$         & 1\\
\ion{Sr}{ii} & $-3.58\pm 0.04$ & 2\\
\ion{Ba}{ii} & $-3.22$         & 1\\
\noalign{\smallskip}\hline\noalign{\smallskip}
\end{tabular}
\end{center}
\end{table}

%%%%%%%%%%%%%%%%%%%%%%%%%%%%%%%%%%%%%%%%%%%%%%%%%%%%%%%%%%%%%%%%%%
\subsection{SDSS\,J133718+074536}

Very few lines are detectable in the spectrum of this ``hot'' EMP star
for a combination of ``high'' temperature, low metallicity, and low
S/N.  For these same reasons, the analysis based on the SDSS spectrum
relies basically on the K-\ion{Ca}{ii} line at 393.3\,nm.  The
continuum determination in this range of the SDSS spectrum is subjective and not
trivial, and the calcium line itself has a
strange shape, which, at the SDSS spectral resolution and S/N, is compatible with both
an extremely low Ca abundance or a damaged line.  These findings can
explain the large disagreement (--0.91\,dex) in the abundances
derived from the analyses of the SDSS and X-Shooter
spectra.

The X-Shooter observation has a relatively short exposure 
time because of a change in weather conditions, thus its S/N
is lower than desired.  Only nine \ion{Fe}{i} lines, six \ion{Mg}{i} lines,
and five Ca lines were detected.

As for the other stars, we do not consider the bluest line of the
\ion{Mg}{i}-b triplet, because its abundance is too high with respect
to the one from the other \ion{Mg}{i} lines. In addition the line at
382.9\,nm is not included in the abundance determination, leading to a far
too high abundance. It was also impossible to reproduce
satisfactorily the line profile by synthesis, owing to its placement 
between two H lines.
The \ion{Ca}{i} line at 422.6\,nm gives an abundance ([Ca/H]=--3.40) in
reasonable agreement with the one from the \ion{Ca}{ii}-K line,
[Ca/H]=--3.23.  The abundance of Ca derived from the IR triplet is
much higher, $-2.87\pm 0.29$, but the region is badly affected by the
sky subtraction.  For the $\alpha$ enhancement determination, we take into account
the Mg lines included in the sample for the abundance determination,
the 422.6\,nm \ion{Ca}{i} line, and the \ion{Ca}{ii}-K line, giving
[$\alpha$/H]=$-3.27\pm 0.14$.  The individual abundances can be found
in Table\,\ref{ab133718}.  No spectral line in the wavelength range of
the Li doublet at 670.7\,nm is clearly visible.

\begin{table}
\caption{\label{ab133718}
SDSS\,J133718+074536
abundances.
}
\begin{center}
\begin{tabular}{llr}
\hline\noalign{\smallskip}
Element & [X/H]$_{\rm fit}$ &  N\\
       & x=2.0\kms         &   \\
\hline\noalign{\smallskip}
\ion{Fe}{i}  & $-3.49\pm 0.32$ & 9\\
\ion{Mg}{i}  & $-3.24\pm 0.19$ & 4\\
\ion{Ca}{i}  & $-3.40 $        & 1\\
\ion{Ca}{ii} & $-2.96\pm 0.30$ & 4\\
\noalign{\smallskip}\hline\noalign{\smallskip}
\end{tabular}
\end{center}
\end{table}

%%% FIGURE %%%%%%%%%%%%%%%%
\begin{figure}
\begin{center}
\resizebox{\hsize}{!}{\includegraphics[clip=true]
{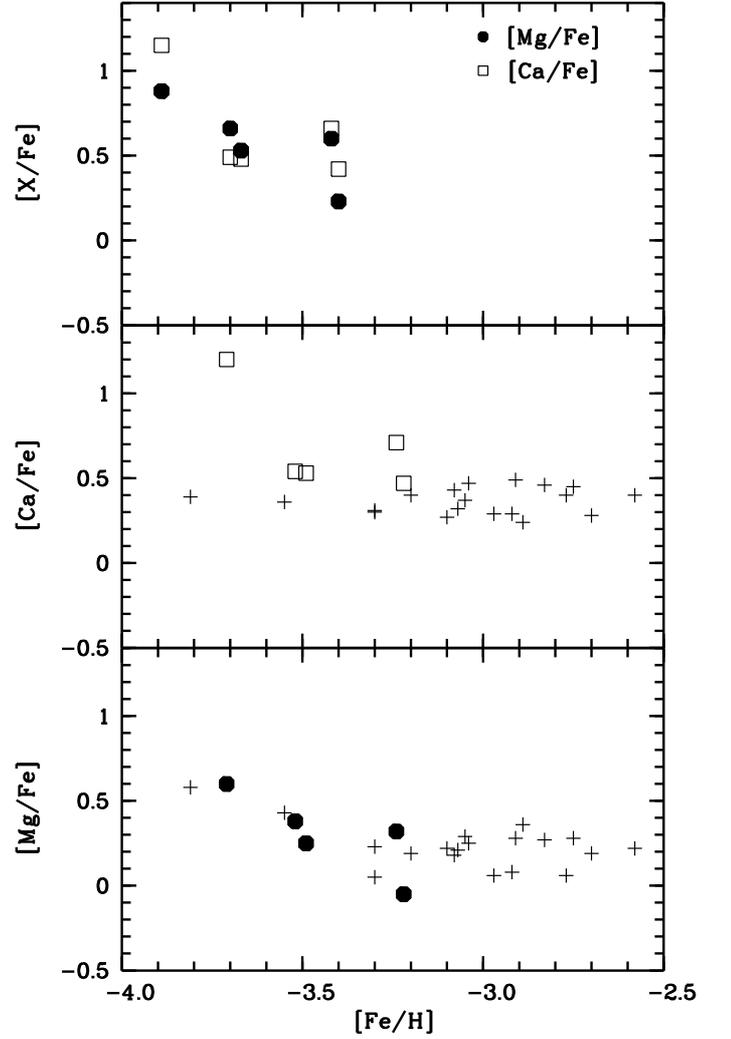}}
\end{center}
\caption[]{In the upper panel, 
the abundance of Mg (filled hexagons) and Ca (square) versus [Fe/H].
To derive [Ca/Fe], the lines of \ion{Ca}{ii} have been taken into account.
The 3D corrections are applied.
In middle and lower panel, the 1D-LTE abundances ration [Ca/Fe] and
[Mg/Fe], respectively, with the same symbols, are compared to the ``First Stars'' results
\citep{bonifacio09}.
}
\label{plotalpha}
\end{figure}
%%% FIGURE %%%%%%%%%%%%%%%%

\section{Kinematics and distance}

All the stars have a measured radial velocity above $\simeq 200$km/s except for
SDSS\,J082511+163459, which has a radial velocity of only 24 km/s.
This indicates that at least the four high velocity stars could be members
of the Galactic halo. We have verified the compatibility of the radial
velocity measured with what is expected from the Besan\c con kinematic
model of the Galaxy \citep{Robin}. For each star, we simulated a
Besan\c con field of 20 sq deg in the SDSS bands centred on the Galactic
coordinates of the given star and then extracted a sample of simulated
stars with magnitudes and colors similar to those of the target star. In this
way, we checked the radial velocity compatibility with the expected
distribution. Owing to the emerging evidence of a Galactic halo populated with
substructures \citep{Belokurov06},
we also explored the possible connection of the
stars with some of them, looking for correspondences in distance and in projection on the
sky.

We also estimated the distance of each target star by fitting five or eight
bands photometry (as given in table \ref{allstar}) with \citet{chieffi03}
metal-poor isochrones, after correcting for interstellar
absorption. The age of the isochrone was fixed at 12.5\,Gyr for all the
stars and the best-fit model was chosen using a $\chi ^2$ technique.
This is a first rough estimate, which will be refined when a more
appropriate set of isochrones (at a compatible low metallicity) become 
available.

\subsection{SDSS\,J044638-065528}

This star has a radial velocity of 242 \kms. With the Besan\c con model,
we have been unable to recover any radial velocity that is so high. We
extracted 40 stars within 0.05\,mag in terms of color and 0.1\,mag in g
magnitude around the star. The calculated average ${\rm V_{\rm rad}}$ is 43.2\kms\
with a dispersion of $\simeq 50.1$\kms. This object is indeed at
about four times the velocity dispersion of the stars in this magnitude
range but still at about 2.5 times the velocity dispersion of the Galactic halo.
The estimated distance modulus is (m-M)$_0$=12.90, which corresponds to
a distance of 3800\,pc. The object does not fall within any known Galactic halo substructure. 
Moreover, the relatively short distance makes this object an unlikely member of
any stream or overdensity.

\subsection{SDSS\,J082511+163459}

This object appears to be somewhat more evolved than the remainder of the sample 
owing to its redder colors. The distance estimate of this star gives a value of 10.9\,kpc
or a distance modulus of (m-M)$_0$=15.19. This value places the object above
the edge of the Galactic disc in the Monoceros Ring region. A velocity of 
only 24\kms\ implies that this object is a possible member of this structure.

\subsection{SDSS\,J085211+033945}

We found that 2\% of the objects close to this star in the Besan\c con
simulation have a ${\rm V_{\rm rad}}$ of $\simeq 228\pm20$\kms. This
star is again located on the border of the Monoceros Ring but
at half the distance of the previous one, the estimated distance modulus being
(m-M)$_0$=13.75 or 5.6\,kpc. 

\subsection{SDSS\,J090733+024608}

This object and the previous one, SDSS\,J085211+033945, are close on
the sky, within less than six degrees of each other. Distances are also similar, 
SDSS\,J090733+024608 being at
4.7\,kpc (distance modulus (m-M)$_0$=13.37). If they also
shared the same ${\rm V_{\rm rad}}$, we could try to associate them with a similar
feature in the Galactic halo. The measured ${\rm V_{\rm rad}}$ is equal to 304 \kms, which, if
the two objects were along a stream, would imply a strong kinematical
gradient that has never been observed. If a more precise ${\rm V}_{\rm rad}$ were measured
for both objects, we still might be able to associate the two
stars.

\subsection{SDSS\,J133718+074536}

This star is located toward the the Sagittarius Tidal Arm, which
encompass the entire north Galactic cap \citep{Belokurov06} at a distance
of $\simeq 10.6$\,kpc. This is far nearer than the tidal arm, which is
located in this region at $\simeq 40$\,kpc. A measured ${\rm V}_{\rm rad}$ of 212\kms\
implies that this star is most likely to be a member of the Galactic halo.

\section{Discussion}

\subsection{Metallicity distribution function}

\begin{figure}
\begin{center}
\resizebox{\hsize}{!}{\includegraphics[draft = \draftflag,clip=true]
{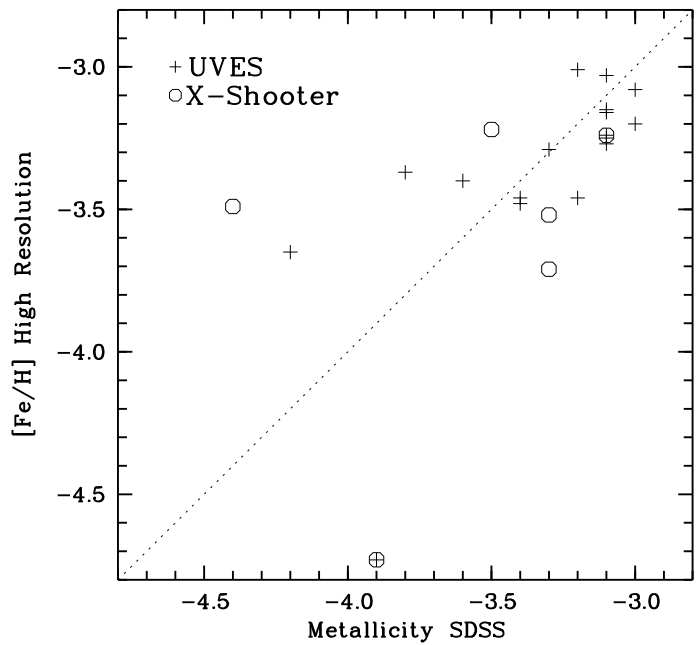}}
\end{center}
\caption[]{The comparison between the metallicity derived from
the low resolution SDSS spectra and our high resolution
observations. 
Crosses represent UVES data \citep{sdss-uves}, and circles
the X-Shooter results. The circle-cross symbol is SDSS\,J102915+172927, 
discussed in \citet{caffau11}, for which we have both X-Shooter
and UVES data. 
}
\label{plot_metcomp}
\end{figure}

The results of this observational campaign show that our method for
selecting EMP stars from the SDSS spectra is robust. The method relies
heavily on the \ion{Ca}{ii} K line and has folded-in an assumed
$\alpha$ element enhancement [Ca/Fe]$=+0.4$.  For stars 
characterised by a low $\alpha$ to iron ratio, as found in
\citet{bonifacio11}, the method underestimates the metallicity.  
The \ion{Ca}{ii} K is such a strong line that it is, sometimes,
the single detectable feature in a low resolution spectrum
of an EMP star. For this reason, several surveys searching for
EMP candidates are based on the stellar ``metallicity'' 
determination from this line, i.e. the
HK survey \citep{beers85,beers92} and the
Hamburg-ESO survey \citep{reimers,christlieb08}.
In Fig.\,\ref{plot_metcomp}, the measurements of [Fe/H], derived in this
work and from UVES spectra in \citet{sdss-uves}, is compared to the
``metallicity'' derived from the SDSS spectra. From the plot, it is
clear that down to [Fe/H]$>-3.5$ the agreement is very good, while, at
lower metallicities, the metallic lines in the SDSS spectra are very
weak, making the low resolution analysis more uncertain.  The good
agreement between the metallicities estimated from SDSS spectra and
the high resolution analysis, shown in Fig.\,\ref{plot_metcomp},
suggests that the metallicities based on the SDSS spectra can be used
to trace the low-metallicity tail of the Galactic halo metallicity distribution
function.  A matter of concern is that below [Fe/H]=--3.5
there is large scatter.
In spite of this, we do not conclude that
our analysis of the SDSS spectra systematically underestimates the metallicity.
By looking at Fig.\,\ref{plot_metcomp},
we see that
one star analysed in this paper (SDSS\,J133718+074536)
and another analysed using UVES spectra \citep{sdss-uves}
have a [Fe/H] that is higher 
by about 1\,dex, than the estimate based on the SDSS spectrum.
However, a third star \citep[shown with a circle-cross symbol][]{caffau11},
for which we have both X-Shooter and UVES spectra,
has a [Fe/H] that is about 1\,dex {\em lower}. 
We infer that for [Fe/H]$\lesssim -3.5$ 
the uncertainty in the metallicity
derived from SDSS spectra is about 1\,dex, but unbiased.

%%% FIGURE %%%%%%%%%%%%%%%%
%%% FIGURE %%%%%%%%%%%%%%%%

A more robust assessment of the systematics between metallicity estimates
derived from SDSS spectra and those based on higher resolution spectra
would require the observation
of a larger set of stars.  The efficiency of X-Shooter suggests that
at a rate of 6 stars/night it should be possible to observe about 100
stars in 17 nights. Such an effort, coupled with some higher
resolution observations (e.g. with UVES) for the most metal-poor
stars, which have hardly any detectable iron lines in the X-Shooter spectra,
would allow us to conclusively confirm the almost vertical drop in the
metallicity distribution function at [Fe/H]$\sim$--3.5 found for the
Hamburg-ESO survey \citep{reimers,christlieb08} by \citet{schork}.
The metallicity distribution function ``as observed'' for our sample
of $\sim$ 125\,000 stars does not display this vertical drop
\citep{bonifacio_rio}.  If the underlying metallicity distribution
function is similar to that of \citet{schork}, we expect that the
observations with X-Shooter of the 100 most metal-poor stars,
according to the SDSS-based estimates, will show that in reality they
are of metallicity above [Fe/H]$=-3.5$.

To reconstruct the ``true'' metallicity distribution function
of the Galactic halo it is necessary to correct for the bias present in the
SDSS sample.  One could argue that there should be no bias, at least
in the low metallicity tail, since the $ugriz$ colours become very
insensitive to metallicity below [Fe/H]$=-2.0$.  \citet{bonifacio_rio}
pointed out, by using a photometric metallicity estimate, that the
sample of SDSS stars with spectra appeared to have a larger fraction
of very metal-poor stars than a random sample, taken from the SDSS
database with the same colours. We can obtain an indication of what is
boosting the number of metal-poor stars in the sample of stars with
SDSS spectra by examining the {\tt Object Type} keyword of the SDSS
spectra of our sample of stars.  This keyword contains information on
the reason for which the star was selected for SDSS spectroscopy.
From Table\,\ref{logbook}, we see that our targets were observed as either
spectrophotometric standards (used for the flux calibration of the
spectra), reddening standards, or ``serendipity blue''. The sample
observed with UVES by \citet{sdss-uves} shows the same descriptors
and, in addition a QSO (i.e. the star was expected to be a QSO from
its colours).  All these classes of objects contain a considerable
fraction of very metal-poor stars, thus boosting the numbers in the
SDSS spectroscopic database.

\subsection{Carbon enhancement}

We note that none of the stars observed shows
any significant enhancement in carbon. The stars were selected as
non-carbon-enhanced from the SDSS spectra. \citet{behara} demonstrated
that it is possible to select carbon-enhanced-metal-poor (CEMP)
from the SDSS spectra. However, one might have expected that at the
higher spectral resolution and S/N of the X-Shooter spectra,
some stars classified as non-carbon enhanced based on the SDSS
spectra, could display a moderate carbon enhancement. Such star has
been found, in contrast to some other studies \citep{beers2006}.

\subsection{$\alpha$ vs. iron abundance}

In Fig.\,\ref{plotalpha}, we show the ratio of either Mg and Ca to Fe, as a
function of iron abundance. 
The same 3D corrections, derived for the SDSS\,J044638-065528 star
from the 6270\,K/4.0/-3.0 model, are applied to all stars.
Three stars (SDSS\,J044638-065528, SDSS\,J085211+163459, 
and SDSS\,J133718+074536) have parameters
compatible with the 3D computations. SDSS\,J090733+024608 is cooler, hence the 3D corrections
should be smaller because of the temperature structure of the 3D model, but the
microturbulence is 0.5\kms\ smaller increasing the 3D corrections.
The two effects, at the precision of the present work, cancel out.

The general picture is compatible with
what is observed for the sample of EMP stars of
\citet{bonifacio09} (see middle and lower panels). 
Star SDSS\,J044638-065528 displays a rather high
$\alpha$ to iron ratio, which appears to be driven by the high Ca abundance.
In the lower panel of Fig.\,\ref{plotalpha}, we show that 3D
corrections are not negligible and provide a closer
consistency between the Ca and Mg abundances.  The analysis was
performed in LTE, although we know that there are indeed NLTE effects
on both Mg \citep{andrievsky} and Ca \citep{mashonkina}. We defer
the NLTE analysis to a future paper, but expect this will achieve
even closer consistency between Ca and Mg and probably a smaller
scatter in the [X/Fe] ratios between different stars, without changing
the general pattern.

\subsection{Sr and Ba abundances}

We find a low [Sr/Ba] ratio of --0.36 for SDSS\,J090733+024608, which
is rather uncommon for metal-poor stars. In the sample of unevolved EMP
stars of \citet{bonifacio09}, none shows such a low Sr/Ba ratio.
However, in the sample of giants of \citet{francois}, BD --18$^\circ$
5550 ([Fe/H]=--3.06) has [Sr/Ba]=--0.27 (see also \citealt{andr2011}).

\balance
\section{Conclusions}

We have confirmed the good performance of the X-Shooter IFU \citep{IFU},
which has been used
as an image-slicer, to obtain medium-resolution spectra suitable for
abundance analysis.  This coupled with the high throughput of
X-Shooter has allowed us to investigate several EMP candidate stars in a
relatively short time.

The sample of faint distant stars that we have selected as candidate
EMP from the SDSS survey has been confirmed to be of extremely low
metallicity. Our selection method based on SDSS spectra has been
proven to be highly reliable. The chemical composition of our sample
appears to be compatible with that found for the much brighter sample
of \citet{bonifacio09}. This may suggest that the stars in the Galactic halo are well mixed
throughout its extension.  We did not find any $\alpha$-poor stars
similar to those found by \citet{bonifacio11}.  The observations suggest
a variability of Li abundances, as already noted in
\citet{sbordone} and an absence of a general carbon enhancement among
EMP stars.

The observation of a large sample of these candidates extracted from
the SDSS is feasible in fewer than 20 nights of observations and would
allow us to understand the shape of the metal-weak tail of the Galactic halo
metallicity distribution function.

%%%%%%%%%%%%%%%%%%%%%%%%%%%%%%%%%%%%%%%%%%%%%%%%%%%%%%%%%%%%%%%%%%%

\begin{acknowledgements}
We acknowledge support from the Programme Nationale
de Physique Stellaire (PNPS) and the Programme Nationale
de Cosmologie et Galaxies (PNCG) of the Institut Nationale de Sciences
de l'Universe of CNRS.
\end{acknowledgements}

%%%%%%%%%%%% APPENDIX %%%%%%%%%%%%%%%%%%%%%%%%%%%%

%
%  A&A article: Metal-poor stars
%
%
%%%%%%%%%%%%%%%%%%%%%%%%%%%%%%%%%%%%%%%%%%%%%%%%%%%%%%%%%%%%%%%%%%%%%%%%%%%
%\appendix
%%%%%%%%%%%%%%%%%% END APPENDIX %%%%%%%%%%%%%%%%%%%%%%%%%%%%%%%

\bibliographystyle{aa}

\begin{thebibliography}{}


\bibitem[Andrievsky et 
al.(2010)]{andrievsky} Andrievsky, S.~M., Spite, M., 
Korotin, S.~A., Spite, F., Bonifacio, P., 
Cayrel, R., Fran{\c c}ois, P., \& Hill, V.\ 2010, \aap, 509, A88 

\bibitem[Andrievsky et 
al.(2011)]{andr2011} Andrievsky, S.~M., Spite, F., Korotin, S.~A., Fran{\c c}ois, P., Spite, M., Bonifacio, P., Cayrel, R., \& Hill, V.\ 2011, \aap, 530, A105 

\bibitem[Barklem et
al.(2000b)]{barklem00b} Barklem, P.~S., Piskunov, N., \& O'Mara, B.~J.\ 2000b, \aap, 363, 1091

\bibitem[Barklem et
al.(2000)]{barklem00} Barklem, P.~S., Piskunov, N., \& O'Mara, B.~J.\ 2000, \aap, 355, L5

\bibitem[Beers(1999)]{beers99} Beers, T.~C.\ 1999, \apss, 265, 547

\bibitem[Beers et al.(1985)]{beers85} Beers, T.~C., Preston, 
G.~W., \& Shectman, S.~A.\ 1985, \aj, 90, 2089

\bibitem[Beers et al.(1992)]{beers92} Beers, T.~C., Preston, 
G.~W., \& Shectman, S.~A.\ 1992, \aj, 103, 1987 

\bibitem[Beers et al.(2006)]{beers2006} Beers, T.~C., Lucatello, 
S., Marsteller, B., Sivarani, T., Barklem, P., Christlieb, N., 
\& Rossi, S.\ 2006, International Symposium on Nuclear Astrophysics - Nuclei in the Cosmos,  

\bibitem[Behara et 
al.(2010)]{behara} Behara, N.~T., Bonifacio, P., 
Ludwig, H.-G., Sbordone, L., Gonz{\'a}lez Hern{\'a}ndez, J.~I., \& Caffau, E.\ 2010, \aap, 513, A72 

\bibitem[Belokurov et al.(2006)]{Belokurov06} Belokurov, V., et 
al.\ 2006, \apjl, 642, L137 


\bibitem[Bessell 
\& Norris(1984)]{BN84} Bessell, M.~S., \& Norris, J.\ 1984, \apj, 285, 622 

\bibitem[Bidelman 
\& MacConnell(1973)]{BM73} Bidelman, W.~P., \& MacConnell, D.~J.\ 1973, \aj, 78, 687 

\bibitem[Bond(1970)]{Bond70} Bond, H.~E.\ 1970, \apjs, 22, 117 

\bibitem[Bonifacio 
\& Caffau(2003)]{bonifacio03} Bonifacio, P., \& Caffau, E.\ 2003, \aap, 399, 1183

\bibitem[Bonifacio et 
al.(2009)]{bonifacio09} Bonifacio, P., et al.\ 2009, \aap, 501, 519 

\bibitem[Bonifacio et al. (2010)]{bonifacio_rio} Bonifacio, P., 
Caffau E., Ludwig, H.-G., Sbordone, L., Gonz\'alez~Hern\'andez, J.I.,
Behara, N. T., 2009, IAU XVII General Assembly, Joint Discussion 5, ed.
J. Binney

\bibitem[Bonifacio et al.(2011a)]{bonifacio11} Bonifacio, P., et 
al.\ 2011, Astronomische Nachrichten, 332, 251 

\bibitem[Bonifacio et al.(2011b)]{sdss-uves} Bonifacio, P., et
al.\ 2011, in preparation

\bibitem[Caffau 
\& Ludwig(2007)]{cl07} Caffau, E., \& Ludwig, H.-G.\ 2007, \aap, 467, L11 

\bibitem[Caffau et al.(2011a)]{abbosun} Caffau, E., Ludwig, 
H.-G., Steffen, M., Freytag, B., \& Bonifacio, P.\ 2011a, \solphys, 268, 255

\bibitem[Caffau et al.(2011b)]{caffau11} Caffau, E., et al.\ 2011b,  Nature, 477, 67

\bibitem[Castelli 
\& Kurucz(2003)]{castelli03}
Castelli, F., \& Kurucz, R.~L.\ 2003, Modelling of Stellar Atmospheres, 210, 20P 
\bibitem[Cayrel et 
al.(2004)]{cayrel04} Cayrel, R., et al.\ 2004, \aap, 416, 1117

\bibitem[Chieffi et al.(2003)]{chieffi03} Chieffi, A., 
Dom{\'{\i}}nguez, I., H{\"o}flich, P., Limongi, M., 
\& Straniero, O.\ 2003, \mnras, 345, 111

\bibitem[Christlieb et al.(2002)]{christlieb} Christlieb, N., et 
al.\ 2002, \nat, 419, 904 

\bibitem[Christlieb et 
al.(2008)]{christlieb08} Christlieb, N., Sch{\"o}rck, T., Frebel, A., 
Beers, T.~C., Wisotzki, L., \& Reimers, D.\ 2008, \aap, 484, 721

\bibitem[D'Odorico et al.(2006)]{dodorico} D'Odorico, S., et 
al.\ 2006, \procspie, 6269E,  98

\bibitem[Edvardsson et al.(1993)]{edvardsson93} 
Edvardsson, B., Andersen, J., Gustafsson, B., Lambert, D.~L., Nissen, P.~E., 
\& Tomkin, J.\ 1993, \aap, 275, 101

\bibitem[Fran{\c c}ois et 
al.(2007)]{francois} Fran{\c c}ois, P., et al.\ 2007, \aap, 476, 935 

\bibitem[Frebel et al.(2005)]{frebel} Frebel, A., et al.\ 
2005, \nat, 434, 871 

\bibitem[Freytag, Steffen, \&
Dorch(2002)]{freytag02}{Freytag}, B., {Steffen}, M., \& {Dorch}, B. 2002,
Astronomische Nachrichten, 323, 213

\bibitem[Freytag et al. (2011)]{freytag10}
Freytag, B. et al.\ 2011, ``Realistic simulations of stellar convection'',
Journal of Computational Physics: special topical issue on computational plasma physics,
ed. Barry Koren

\bibitem[Guinouard et al.(2006)]{IFU} Guinouard, I. et al.\ 2006, 
\procspie, 6273E, 116 

\bibitem[Goldoni et al.(2006)]{goldoni} Goldoni, P., Royer, F., 
Fran{\c c}ois, P., Horrobin, M., Blanc, G., Vernet, J., Modigliani, A., 
\& Larsen, J.\ 2006, \procspie, 6269, 80  

\bibitem[{{Kelson}(2003)}]{kelson03} 
Kelson, D.D., PASP, 115, 688

\bibitem[{{Kurucz}(1993)}]{kurucz93}
{Kurucz}, R. 1993, SYNTHE Spectrum Synthesis Programs and Line
Data.~Kurucz CD-ROM No.~18.~Cambridge, Mass.: Smithsonian Astrophysical
Observatory, 1993., 18

\bibitem[{{Kurucz}(2005)}]{kurucz05}
{Kurucz}, R.~L. 2005, Memorie della Societ\`a Astronomica
 Italiana Supplementi, 8, 14

\bibitem[Lodders et al.(2009)]{lodders09} Lodders, K., Plame, H., 
\& Gail, H.-P.\ 2009, Landolt-B{\"o}rnstein - Group VI Astronomy and Astrophysics
Numerical Data and Functional Relationships in Science and Technology Volume 
4B: Solar System.~ Edited by J.E.~Tr{\"u}mper, 2009, 4.4., 44 

\bibitem[Ludwig et al.(2008)]{ludwig08} Ludwig, H.-G., 
Bonifacio, P., Caffau, E., Behara, N.~T., Gonz{\'a}lez Hern{\'a}ndez, 
J.~I., \& Sbordone, L.\ 2008, Physica Scripta Volume T, 133, 014037

\bibitem[Ludwig et al.(2009)]{ludwig09} Ludwig, H.-G., Caffau, 
E., Steffen, M., Freytag, B., Bonifacio, P., 
\& Ku{\v c}inskas, A.\ 2009, \memsai, 80, 711

\bibitem[Mashonkina et 
al.(2007)]{mashonkina} Mashonkina, L., 
Korn, A.~J., \& Przybilla, N.\ 2007, \aap, 461, 261 

\bibitem[Molaro 
\& Castelli(1990)]{molaro90} Molaro, P., \& Castelli, F.\ 1990, \aap, 228, 426

\bibitem[Molaro 
\& Bonifacio(1990)]{MB90} Molaro, P., \& 
Bonifacio, P.\ 1990, \aap, 236, L5 

\bibitem[Norris et al.(2007)]{norris} Norris, J.~E., 
Christlieb, N., Korn, A.~J., Eriksson, K., Bessell, M.~S., Beers, T.~C., 
Wisotzki, L., \& Reimers, D.\ 2007, \apj, 670, 774 

\bibitem[Qian 
\& Wasserburg(2008)]{QW} Qian, Y.-Z., \& Wasserburg, G.~J.\ 2008, \apj, 687, 272 

\bibitem[Reimers(1990)]{reimers} Reimers, D.\ 1990, The 
Messenger, 60, 13 

\bibitem[Robin et 
al.(2003)]{Robin} Robin, A.~C., Reyl{\'e}, C., Derri{\`e}re, S., \& Picaud, S.\ 2003, \aap, 409, 523

\bibitem[{{Sbordone}(2005)}]{sbordone05}
{Sbordone}, L. 2005, Memorie della Societ\`a Astronomica Italiana Supplementi, 8, 61

\bibitem[{{Sbordone} {et~al.}(2004){Sbordone}, {Bonifacio}, {Castelli}, \&
 {Kurucz}}]{sbordone04}
{Sbordone}, L., {Bonifacio}, P., {Castelli}, F., \& {Kurucz}, R.~L. 2004,
 Memorie della Societ\`a Astronomica Italiana Supplementi, 5, 93

\bibitem[Sbordone et 
al.(2010)]{sbordone} Sbordone, L., et al.\ 2010, \aap, 522, A26 

\bibitem[Sch{\"o}rck et 
al.(2009)]{schork} Sch{\"o}rck, T., et al.\ 2009, \aap, 507, 817 

\bibitem[Slettebak 
\& Brundage(1971)]{SB71} Slettebak, A., \& Brundage, R.~K.\ 1971, \aj, 76, 338 

\bibitem[Stehl{\'e} 
\& Hutcheon(1999)]{stehle99} Stehl{\'e}, C., \& Hutcheon, R.\ 1999, \aaps, 140, 93 

\bibitem[{{Van Dokkum}(2001)}]{vandokkum01}
Van Dokkum, P.G., 2001, PASP, 113, 1420

\bibitem[York et al.(2000)]{sdss} York, D.~G., et al.\ 2000, 
\aj, 120, 1579

\end{thebibliography}

\end{document}